# Accelerating Cavity Fault Prediction Using Deep Learning at Jefferson Laboratory

**Md. M. Rahman**[1], **A. Carpenter**[2], **K. Iftekharuddin**[1], **C. Tennant**[2]

[1]Old Dominion University, Norfolk, VA 23529
[2]Thomas Jefferson National Accelerator Facility, Newport News, VA 23606

E-mail: mrahm006@odu.edu

## Abstract

Accelerating cavities are an integral part of the Continuous Electron Beam Accelerator Facility (CEBAF) at Jefferson Laboratory. When any of the over 400 cavities in CEBAF experiences a fault, it disrupts beam delivery to experimental user halls. In this study, we propose the use of a deep learning model to predict slowly developing cavity faults. By utilizing pre-fault signals, we train a LSTM-CNN binary classifier to distinguish between radio-frequency (RF) signals during normal operation and RF signals indicative of impending faults. We optimize the model by adjusting the fault confidence threshold and implementing a multiple consecutive window criterion to identify fault events, ensuring a low false positive rate. Results obtained from analysis of a real dataset collected from the accelerating cavities simulating a deployed scenario demonstrate the model's ability to identify normal signals with 99.99% accuracy and correctly predict 80% of slowly developing faults. Notably, these achievements were achieved in the context of a highly imbalanced dataset, and fault predictions were made several hundred milliseconds before the onset of the fault. Anticipating faults enables preemptive measures to improve operational efficiency by preventing or mitigating their occurrence.

## 1. Introduction

In many industrial settings, the recent abundance of data from sensors and diagnostics, in conjunction with rapid developments in machine learning, have motivated research in the area of fault prediction. The primary goal of fault prediction is to proactively detect deviations from expected patterns or states, thereby enabling timely intervention to prevent or mitigate potential issues before they escalate into significant problems. Whether the prediction is achieved on longer time scales allowing for preventative maintenance, or on short time scales that require systems to respond quickly, the goal is the same, reduced downtime. In this way, fault prediction contributes to improved operational efficiency and cost savings. In this work, we describe the development of a fault prediction model for a particle accelerator.

### 1.1 Continuous Electron Beam Accelerator Facility (CEBAF)

A particle accelerator is a scientific device used to accelerate charged particles, such as protons or electrons, to very high speeds and energies. These accelerators represent some of the most complex scientific instruments ever designed, built, and operated. They play an instrumental role in advancing our understanding of the fundamental properties of matter and the universe. In addition to particle physics, accelerators also have applications in other fields. They are used in nuclear physics to investigate the structure and forces in the nucleus, in materials science to analyze the structure and properties of materials, in medicine for diagnostics and treatment, radiation therapy, and medical imaging [1].

The Continuous Electron Beam Accelerator Facility (CEBAF) at Jefferson Laboratory is a continuous-wave recirculating linear accelerator (linac) used to deliver electron beams to four experimental halls simultaneously [2-4]. See Fig. 1. CEBAF consists of two antiparallel linacs connected by recirculation arcs. Each linac is comprised of 25 cryomodules and each cryomodule contains 8 superconducting radio-frequency (SRF), or accelerating, cavities. These resonant cavities accelerate electrons, and by making multiple recirculations through the linacs, CEBAF can generate electron energies up to 12 GeV.

Over the years, one of the largest contributors to short machine downtime trips (defined as events which are resolved in less than 5 min) has consistently been caused by SRF system faults. It is important to realize that while the fault is localized to a single cavity – one of many thousands of beamline components – it forces the machine to turn off delivery of beam until the fault is resolved. Experiments often run for several consecutive weeks and the accumulation of these fault events results in a loss of valuable beam time. For example, during a recent operational run from June 2022 through March 2023, SRF cavity trips led to over 320 hours of lost beam time. Further, the time lost in the experimental halls is effectively greater, since data are discarded 30 sec before each trip and 30 sec after every recovery.

Consequently, seeking ways to improve operational efficiency so as to maximize scientific output remains a high priority. In recent years, machine learning (ML) has provided new tools to address this issue, with application to anomaly detection, classification, and prognostics. Given that RF cavities are the fundamental building blocks of particle accelerators and that these devices generate information-rich data, research has been directed toward detection, isolation, classification, and prediction of anomalies in RF systems specifically [5-9].

Recent efforts at Jefferson Laboratory have addressed fault classification, that is, using ML to identify the type of cavity fault [5]. While useful, it represented a post-mortem analysis. Transitioning to a proactive approach, this work explores the use of deep learning to find precursors in RF signals such that we can accurately predict – in a timely manner – if a fault will occur. A fully realized fault prediction system contains two critical components: (1) a robust and accurate fault prediction model, and (2) the implementation of preemptive measures to avoid a predicted fault. The focus of this work is to establish the first step. Such a system would offer the potential of significantly reduce downtime by enabling timely and targeted interventions based on the predictive capabilities of the system.

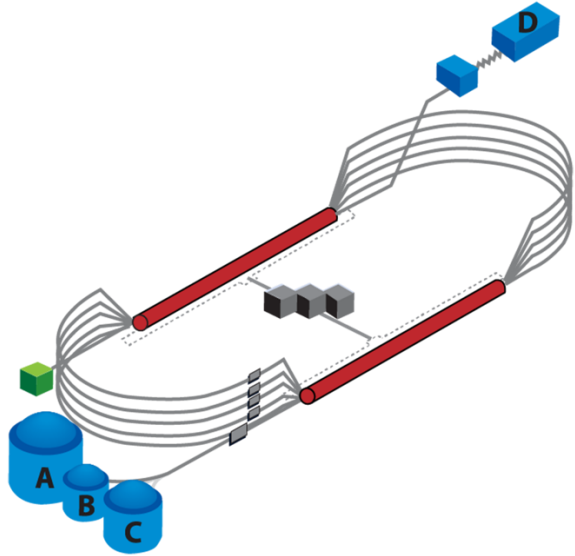

FIGURE 1: Schematic of CEBAF. The two long straight sections (red) represent the two linacs used to accelerate electrons and are connected by recirculating arcs (gray). The experimental end stations are denoted as Halls A, B, C, and D (blue).

## *1.2 Contribution*

In this work, we introduce a deep learning model for fault prediction that leverages multivariate time series data from accelerating cavities. In order to be actionable and effective in practice, the model must be able to forecast impending faults at least 100 ms in advance. Additionally, the system needs to be extremely precise. During a 24-hour period of typical operation, less than 5 faults occur across all eight cavities of a cryomodule, which need to be distinguished from hundreds of thousands of normal presentations of data.

To address these challenges, we make several important contributions. First, we collect and curate data from across several years, using the RF system's built-in data acquisition capability, to effectively train and test a model. See Section 3. Second, we introduce an uncertainty quantification approach, providing a measure of confidence associated with each model prediction. See Section 4.1. Third, to bolster the model's tolerance to noise in the system, we optimize our model through the implementation of a consecutive fault prediction window approach. This strategy aims to improve the robustness of predictions by considering patterns across successive prediction windows. See Section 5.1. Fourth, we describe a method to set a confidence threshold to maintain the necessary precision by minimizing false positives. See Section 5.2. Lastly, absent the ability to access continuously streaming signals, we create a realistic experiment by collecting an imbalanced dataset on which to test and evaluate model performance. See Section 6.

## 2. Related Works

In this section, we briefly highlight some recent fault prediction research with application to particle accelerators (see also Ref. [10] for a more thorough literature review of time series forecasting methods in accelerators). Blokland et al. [11] present a machine learning method that incorporates uncertainty aware predictions, employing the Siamese neural network model to predict upcoming errant beam pulses. The model has the ability to predict impending failures in the accelerator and abort beam pulses before damage occurs. Reščič et al. [12] propose machine learning approaches to predict machine failures via beam current measurements using random forests. In follow-up research, Reščič et al. [13] presents a binary classifier that is capable of predicting accelerator failures with millisecond classification time and a 0% false positive rate. Li et al. [14] propose a forecasting model of the interlocks of a high-intensity proton accelerator. The approach utilizes a Recurrence Plot-Convolutional Neural Network (RPCNN) model for performing the classification task. Lobach et al. [15] present an unsupervised anomaly detection method to predict trips in the magnet power supplies. This work uses an autoencoder network that is trained using normal operation data to identify trip precursors by measuring their level of deviation. Krymova et al. [16] propose to use an autoregressive modeling approach for beam loss prediction at the Large Hadron Collider (LHC) at CERN. This method uses the previous beam loss value in addition to the observed control parameters. Obermair et al. [7] employ a machine learning approach to analyze high-gradient cavity data from CERN's test stand for the Compact Linear Collider (CLIC). The primary objective is to identify previously unrecognized features associated with the occurrence of RF breakdowns. Their work utilizes explainable artificial intelligence (AI) to interpret the parameters of learned models. Radaideh et al. [17] presents an approach of early fault detection in particle accelerator power electronics to reduce their catastrophic failures and improve particle accelerator reliability.

## 3. Data

Fundamental to cavity fault prediction is developing a model which takes as input a specified length of RF signals and can distinguish between normal, stable operation and signals that portend a fault. In order for a model to learn those distinctions, a dataset containing a sufficient number of stable examples and examples of pre-fault data must be collected. Because at present the CEBAF SRF cavities cannot continuously stream data, we must rely on periodic sampling from the RF system's built-in data acquisition system to build up a realistic, representative dataset. The system can be operated in two different modes: one for collecting fault data and the other for collecting normal data. In the both cases the data are 8,192-point waveforms sampled at 5 kHz, yielding 1,638.4 ms snapshots. We utilize four specific signals from each cavity: the measured gradient (GMES) in MV/m, requested klystron output (GASK) in volts, cavity forward RF power (CRFP) in kW, and cavity detune angle (DETA2) in degrees. For this work, the faulty and normal running data are collected from a single cryomodule and the focus is on building a cryomodule-specific model.

## 3.1. Fault Data

In order to capture waveforms of fault events, an acquisition system has been implemented which is triggered by a fault onset. The system utilizes a buffer so that when data is written to file, the waveform not only includes the fault event, but pre- and post-fault data was well. Specifically, the system is configured such that $t = 0$ corresponds to the fault onset. The 1,536 ms of pre-fault data is used to train a model for fault prediction, whereas the post-fault data ($t > 0$) is neglected. An example of the four RF signals during a fault event is shown in Fig. 2. The horizontal axis is time (in milliseconds), and the vertical axis is the magnitude of each signal.

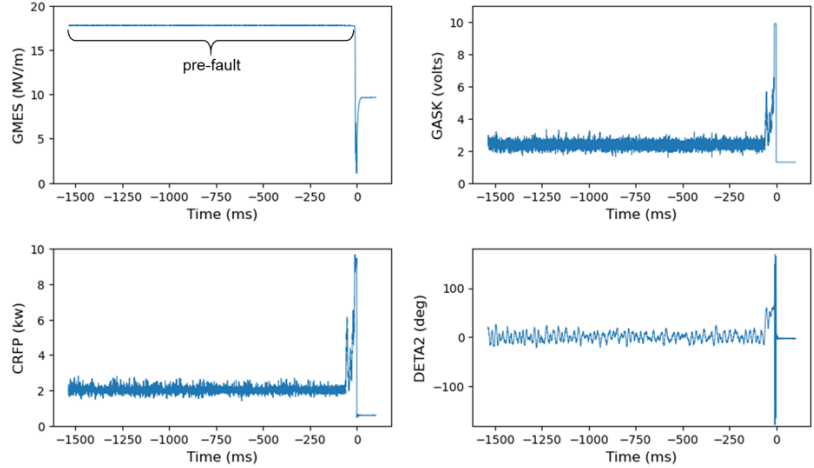

FIGURE 2: Plot of the four RF signals (GMES, GASK, CRFP, DETA2) from a cavity fault event containing pre-fault, fault onset ($t = 0$), and post-fault data.

It is important to note that there are several different well-defined presentations of fault events. That is, subject matter experts are able to find patterns of behavior such that a fault event can be classified into one of eight types. For more details on these fault types and the use of ML for classifying them, see Ref. [5]. For the purposes of this work, it is only necessary to know that these faults can broadly be categorized into two groups, slow and fast developing. The former typically exhibit discernible indications of anomalous behavior prior to the fault, whereas fast developing faults do not (at least in the four signals that are available for fault analysis in this study). Fast growing faults can occur suddenly and without prior indicators, making them particularly challenging to detect and anticipate.

## 3.2. Normal Data

Rather than being automatically triggered by an event in CEBAF, collecting baseline waveforms during normal operation is achieved by executing a script to fetch the data. At the time of this work, this mode of collection was incompatible with the fault data collection system. Only one system could be operational at a time. Consequently, we collected normal running data during two different dedicated time periods. The first was December 14-15, 2022, and the second was March 3-6, 2023. In each instance a program would repeatedly execute the data collection script. Due to lags in sending and writing data, the datasets are semi-continuous in nature. That is, the 1,638.4 ms signals are collected with an approximate 15 second delay before the next set of waveforms can be collected. An example of the waveforms for normal running cavity is given in Fig. 3. Unlike the fault signals, $t = 0$ does not have special significance, other than to mark the start of the data.

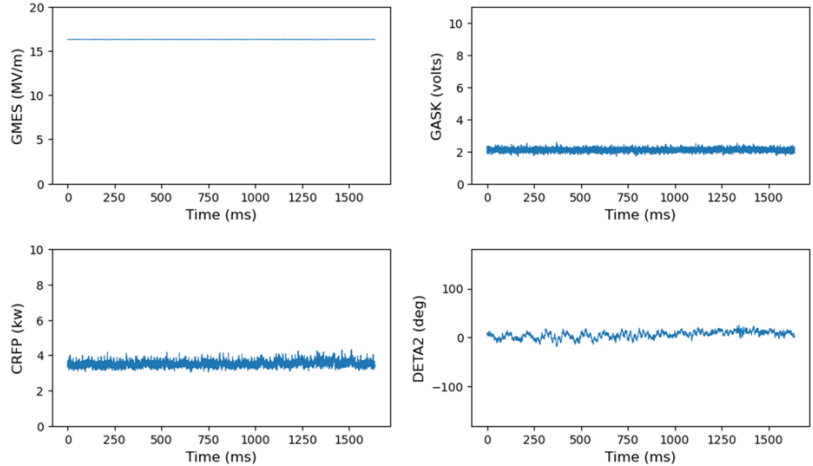

FIGURE 3: Plot of the four RF signals (GMES, GASK, CRFP, DETA2) from a cavity during normal operation.

In order to build up a dataset for model training, we utilize 1,000 normal running events from each of the two collection dates (December 2022 and March 2023), for a total of 2,000 events. Among these events, we assign approximately 80% (1,600 events) for training and reserve the remaining 20% (400 events) for testing. We also incorporate 954 unique, slow-developing fault events, gathered over a span of several years (from Spring 2019 to Fall 2022). We allocate approximately 84% (804 events) for training and reserve about 16% (150 events) for testing. Table 1 shows the distribution of faulty and normal running events for model training and testing.

TABLE 1: Data distribution for model development.

| Data type | # of events | | |
| --- | --- | --- | --- |
| | Training | Test | Total |
| Faulty Events | 804 | 150 | 954 |
| Normal Events | 1600 | 400 | 2000 |

# 4. Model Development

In this section we describe the model architecture, training, and performance metrics.

## 4.1. Model Architecture

To effectively learn from multivariate time series data, we leverage a joint Long Short-Term Memory (LSTM) [18] and Convolutional Neural Network (CNN) [19,20] architecture for our binary classifier model. Figure 4 illustrates the architecture. In the CNN branch, a series of convolutional layers are sequentially applied, with each layer followed by batch normalization and rectified linear unit (ReLU) activation functions. The multiple convolutional layers extract local features from the data. Dropout layers are integrated into the architecture to prevent overfitting by randomly dropping out units during training. Following the convolutional layers, average pooling is employed to downsample the spatial dimensions of the feature maps. Downsampling reduces the number of parameters, lowers computation costs, and helps to prevent overfitting by capturing only the most relevant features. In the LSTM branch, a two-layer LSTM network is employed to capture long-term dependencies of the time series sequence. LSTM cells are used to create a recurrent neural network that can make predictions relating to sequences of data. Dropout layers are used to regularize during training and prevent overfitting. The LSTM branch extracts the temporal features from the data enabling the model to effectively capture the dynamic patterns and dependencies present in the time series data. The CNN and LSTM branches act on the input in parallel. The spatial features produced by the CNN branch and the temporal features generated by the LSTM layers are then concatenated and passed through a fully connected layer.

To offer confidence measures for model predictions, we employ Monte Carlo Dropout (MCDO) a method for uncertainty quantification [21]. Unlike traditional dropout, which is applied solely during training and discarded during inference, MCDO extends dropout to inference (testing or prediction). This involves conducting multiple forward passes through the network with dropout activated and subsequently averaging the predictions. By doing this, MCDO provides a more robust estimate of uncertainty associated with the predictions made by the model.

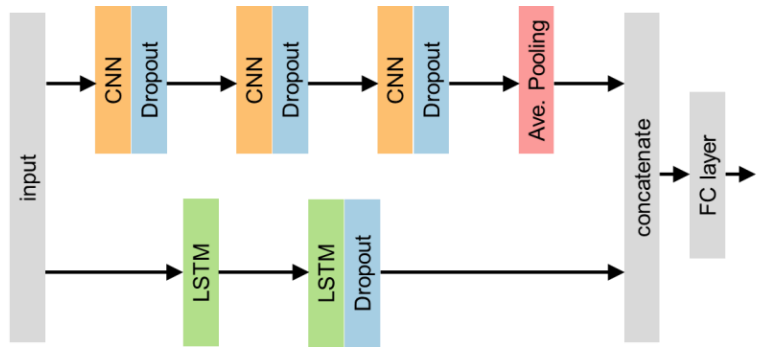

FIGURE 4: Schematic of the joint CNN-LSTM model architecture used for binary classification.

## 4.2. Model Training

Before model training can begin, one critically important decision is the size of the input data. Several considerations constrain our choice. On the one hand, we need enough data points such that meaningful features can be extracted by the model. On the other hand, if the input size is too large, the model will have fewer opportunities to provide a fault prediction before a fault onset. In the end, however, a discussion with RF engineers about what size would be feasible in a future system, considering latencies in moving data through the system, led to the choice of 100 ms (500 data points at our 5 kHz sampling frequency).

Throughout the study, it is useful to split data from the waveforms in a single event (fault or normal) in various ways. When a 100 ms section of data is randomly extracted from the signals, we refer to it as a *sample*. Because they are generated randomly, samples are permitted to overlap. When consecutive 100 ms sections of data are extracted from the signals – without any overlap – each section is referred to as a *window*. Lastly, when the entire 1,638.4 ms signal is considered, it is referred to as an *event*.

An effectively trained model needs to be able to distinguish clearly impending-fault data from clearly normal data. This presents a potential difficulty, however, since fault signatures diminish as one moves further from the onset. That is, regions far from the fault onset ($t = 0$) may appear normal (see Fig. 2). Therefore, during training we only choose pre-fault data where the leading edge of the 100 ms window is in the range from 505 ms to 5 ms prior to the fault onset. For training we select 50 random samples from this region for each of the 804 fault examples. We do not have those constraints for the normal events and select 50 random samples from each of the 1,600 normal training examples. Collectively, this provides 120,200 samples of training data (40,200 fault and 80,000 normal). We employ $z$-score normalization on each channel of the sample signals, ensuring that the mean value of each signal is 0 and the standard deviation is 1. We allocated 80% of these samples for training and the remaining 20% for validation. To assess the model's performance, we withheld a test set comprised of 150 faulty events and 400 normal running events, randomly selected from the total events pool.

## 4.3. Model Performance

Our LSTM-CNN model is developed using the Python [22] programming language and PyTorch [23] deep learning framework. To accelerate the model building process, we utilized the NVIDIA V100 GPU. We employ the Adam optimizer [24] with a learning rate of 0.0001, a batch size of 20, and implemented early stopping with a patience of 20 epochs. We adopt a strategy to save the best model based on its validation performance. Upon completion of the training process, we evaluate the model using the test data.

During the testing phase, we activate dropout layers within both the CNN and LSTM branches. This means that when we pass the same input through the network multiple times (64 times in our case), turning on dropout layers randomly removes connections within the network, effectively creating slightly different architectures for each pass. We then aggregate these outputs to generate our final prediction. This technique helps to enhance the robustness and generalization capability of the model by reducing overfitting and encouraging the network to learn more diverse features from the data. The model's overall accuracy on validation data is 99.99% and 99.26% for the test data. Comparable accuracy between the two data sets provides assurance that the model is not overfitted. While we use randomly generated samples for training and initial model evaluation, it is more insightful to consider model performance on the test data in terms of window, rather than sample, performance. Table 2 displays the confusion matrix when the trained model is applied to the withheld test set. In testing, we use 150 faulty and 400 normal running events which yields 2,250 fault windows and 6,400 normal windows, respectively (because the normal running data does not contain a fault, we can use 16, rather than 15, non-overlapping windows per signal). The results in Table 2 indicate a fault accuracy of 87.82% and normal accuracy of 99.97%. In a deployed system the model is expected to continuously make predictions and minimize false positives – even at the expense of the true positive rate. This is discussed more fully in Section 5.

TABLE 2: Window wise performance of the binary classifier model on the test set.

|  |  | Predicted Label | |
| --- | --- | --- | --- |
|  |  | Normal | Faulty |
| True Label | Normal | 6,398 | 2 |
|  | Faulty | 274 | 1,976 |

In addition to metrics that describe the model's accuracy, for prediction tasks knowing how far in advance a correct prediction can be made is critically important. If the model makes accurate predictions but can only do so while leaving insufficient time to practically implement a mitigating strategy, then such a model is useless. The predictive power of our trained model on the withheld faulty test data is illustrated in Fig. 5, where the model accuracy is evaluated and plotted for each of the 15 windows prior to fault onset. These windows were positioned with their leading edge starting from 5 ms to 1405 ms before the onset of faults, spaced 100 ms apart with no overlap.

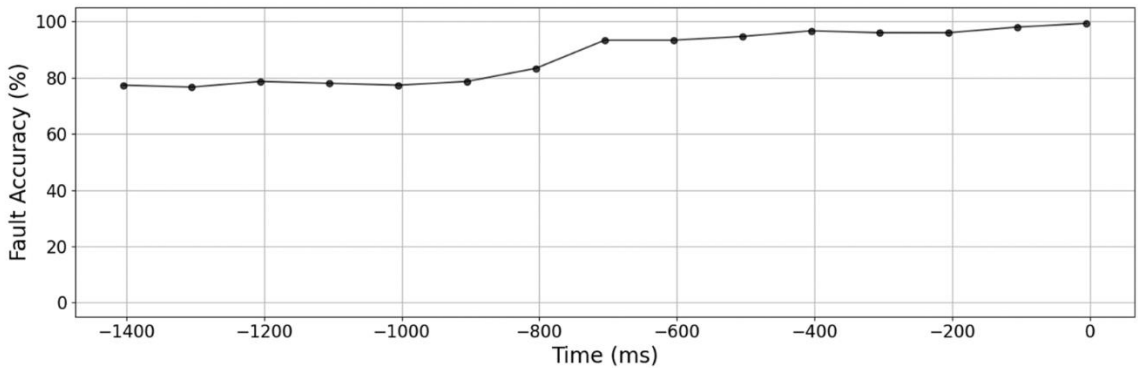

FIGURE 5: The predictive power of the trained model illustrated by plotting the accuracy for the 150 withheld test faults as a function of time prior to onset.

Not surprisingly, as we approach the fault onset ($t = 0$), the accuracy improves. What is especially impressive about the result, however, is that even more than 1 second away, the model is able to accurately predict the onset of a fault with approximately 80% accuracy. That the model can do this having been trained only on data within 505 ms of the fault onset indicates that it has done an excellent job of extracting and learning relevant features from the data.

## 5. Model Optimization

The goal of this work is to use deep learning to extract information from RF signals to accurately predict if a fault will occur. If the model can do so early enough such that a preemptive mitigation strategy can be applied. This opens up exciting possibilities for future cavity control system designs. It is imperative, however, that the model minimize false positives (incorrectly classifying a normal signal as faulty) even if it means sacrificing true positive rate. This is because we want to avoid making any unnecessary changes to the machine during operation. As described in Section 4, the model exhibits excellent predictive power and shows good, but not good enough, performance as a classifier. Consider that if this system was deployed in CEBAF and we had access to continuously streaming RF signals, then in a 24-hour day, the model would make inferences on 864,000 windows of 100 ms data for each cavity in the cryomodule. Based on historical data, any of the eight cavities will fault on average twice per day. Assuming our model identifies the precursors of a fault 1 second in advance, then the model needs to correctly identify 20 windows (2 faults/day × 10 windows/fault) out of 6,912,000 (864,000 windows/cavity × 8 cavities/cryomodule) as faulty, while keeping the false positives to a minimum in a very imbalanced dataset. If we applied the false positive rate from the results in Table 2 to the present example, there would be over 2,000 false positive predictions per day. To address this, there are two parameters that we can optimize; the number of consecutive windows faults used to make a fault prediction and the fault confidence threshold. Descriptions of how each are chosen are detailed below.

## 5.1. Number of Consecutive Windows

The first parameter to optimize is the number of consecutive windows the model should use to make a faulty prediction. By default, if any of the windows in an event are classified as faulty, then conclude that the event is a fault. However, from practical experience it is not uncommon for cavities to experience brief bursts of noise – unrelated to any fault mechanism – that the model may interpret as indicating a fault. We want to avoid a situation of making a prediction too hastily. On the other hand, if we want to be more conservative and impose the constraint that, say, five consecutive windows need to be classified as a fault by the model before it makes a final prediction, then it jeopardizes our ability to make a timely prediction such that a mitigation measure can be executed. For this work, based on the aforementioned heuristic arguments, we arrive at a value of three consecutive windows, though acknowledge that there exists flexibility in the choice of this parameter.

## 5.2. Fault Confidence Threshold

The binary classifier uses SoftMax [25] activation functions in the last fully connected layer and by default, if the output exceeds 0.5, the input is labeled as faulty. By a judicious choice of fault confidence threshold, we can minimize the false positive rate. To do the optimization we propose using the trained model to run inference on an unseen set of normal events and plotting the resulting model accuracy as a function of fault confidence threshold. The threshold is chosen as the fault confidence value that yields the desired false positive rate. The first 125 normal events from each cavity collected during March 3-6 were incorporated into the data for model training and testing. We use the next 1,670 normal events from each cavity as the unseen data on which to optimize the fault confidence threshold. The various data sets and the role they serve is illustrated in Fig. 6.

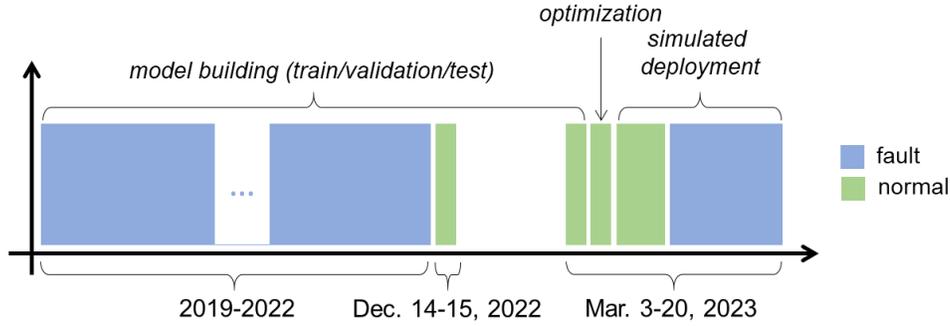

FIGURE 6: A summary of the various fault and normal datasets collected and their use in model development, optimization, and in a simulated deployment.

Because the RF signals and noise profiles vary from cavity to cavity, it is necessary to compute a unique fault confidence threshold for each cavity. Figure 7 displays the results of the accuracy as a function of fault confidence for cavity 2. Note that the model accuracies take into account the constraint from the previous subsection, namely that three consecutive windows must be classified as fault for the event to be predicted as faulty. We repeat this procedure for the remaining seven cavities in the cryomodule. In this way the resulting fault confidence thresholds for cavities 1 through 8 are found to be (0.996, 0.980, 0.700, 0.997, 0.990, 0.500, 0.990, 0.996), respectively.

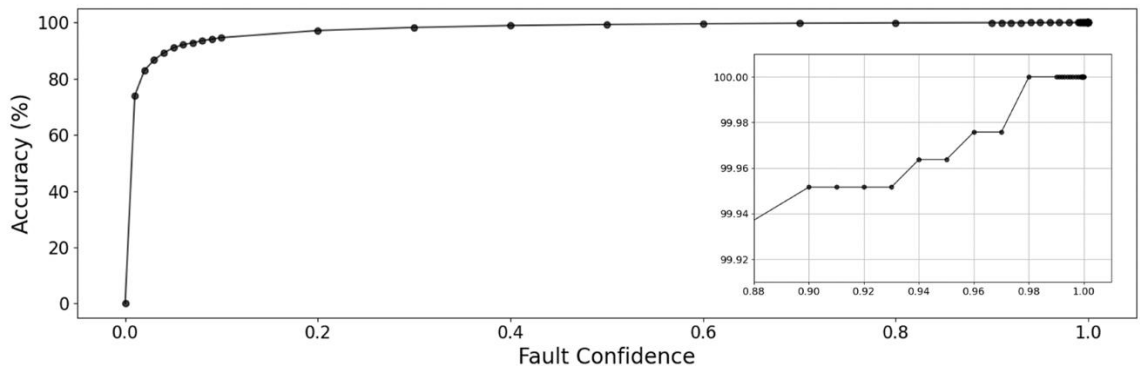

FIGURE 7: Model accuracy on 1,670 normal fault events versus fault confidence for cavity 2. To choose the threshold we find the lowest fault confidence that yields no false positives (100% accuracy). For cavity 2 this threshold is 0.980 (inset).

## 6. Results

With a trained and optimized model, we collected a new dataset, ran inference to simulate deployed performance, and assessed the results. Due to our inability to continuously stream RF signals from cavities, we used the two data acquisition methods described in Section 3 and created a separate dataset intended to mimic what the model would see if deployed in CEBAF. Because the data acquisition systems for collecting normal running and fault events cannot run simultaneously, we combine data from nearby, but different, date ranges. As described in Section 5.2, we leveraged the first 125 normal events from each cavity collected during March 3-6 for model training and the subsequent 1,670 events for model optimization. We now set aside the following 6,500 normal events for the deployed dataset. Because we do not capture any fault events from March 3-6, we use the faults captured in the cryomodule from March 7-20. During that nearly 2-week period, we collected 33 faults which were labeled by a subject matter expert and also added to the deployed dataset. We then use the trained and optimized model to run inference on the newly created deployed dataset. To analyze the results, it will be convenient to discuss the model's performance on the normal and fault events separately.

The performance of the model for each cavity on the 6,500 normal events is summarized in the Table 6. Impressively, the model achieved perfect performance for seven of the eight cavities, while cavity 3 exhibited 99.90% accuracy, with seven misclassified events. Further investigation of the signals associated with these misclassified events revealed that the noise levels were notably higher and persisted for longer durations compared to signals of the preceding and following timestamps. For more details and a representative misclassified event see Appendix A. The takeaway is that, while these seven events were misclassified, it is clear why the model mistook them for portending a fault. We can see this reflected in the fact that the fault confidence threshold for cavity 3 is the relatively low (0.7). It's likely that the 1,670 events used to optimize the fault confidence were especially free from noise for cavity 3, leading to the threshold to be set artificially low. This highlights a major disadvantage of not being able to evaluate the model performance directly on continuously streaming data. If the capabilities existed to deploy the system in CEBAF we could spend 1 hour and analyze the results of the model inference on 360,000 windows and could tune the fault confidence threshold in real-time to achieve the desired false positive rate.

TABLE 6: Deployment performance on 6,500 unique normal events for each cavity.

|  | Cavity | | | | | | | |
|---|---|---|---|---|---|---|---|---|
|  | 1 | 2 | 3 | 4 | 5 | 6 | 7 | 8 |
| Accuracy (%) | 100 | 100 | 99.90 | 100 | 100 | 100 | 100 | 100 |

The model's performance on the 33 fault events is summarized in Table 7. It is important to note that of the 33 faults, 28 are of the fast-developing type and the remaining 5 are slow-developing. We were constrained by how many faults we could collect as March 20 was the last day of CEBAF operation before a several months long scheduled accelerator down for maintenance and upgrade activities. In Section 4 we discussed how model training utilized only pre-fault data from slow-developing faults. Therefore, it is no surprise that the model does so poorly identifying fast faults – it was not trained to do so. On the other hand, we see that the model is able to correctly identify 4 of the 5 slow developing faults.

TABLE 7: Deployment performance on 33 unique fault events.

| Event Type | # of Events | Accuracy (%) |
|---|---|---|
| Slow Faults | 5 | 80 |
| Fast Faults | 28 | 3.57 |

To evaluate how robust the correct faulty predictions are, we offset the start of the first window in 0.2 ms steps 500 times. Because the fault events have onsets all at $t = 0$, the model sees windows at the same set times prior to the fault. By offsetting the start of the initial window (and therefore all subsequent windows), we effectively input 500 different presentations of the pre-fault data to the model. During this exercise we track the time associated with the leading edge of the third consecutive window classified as a fault, which allows us to evaluate the robustness of the predictions as well as quantify the predictive power. For 3 of the 4 correct slow-fault predictions, the model predicted the first three windows as faulty for every one of the 500 offsets. This demonstrates the robustness of the model predictions. The earliest prediction time was 1,205 ms (offset of 0 ms) and the latest was 1,105 ms (offset of 500 ms) prior to the fault onset. To accurately predict even some slow faults with over 1 second advanced notice is a remarkably good result. It should be noted that we are constrained by the 1,638.4 ms length of the events we collect. These results suggest that with streaming data, where we are not limited by the length of the event, it may be possible to predict some slow fault events even farther in advance. When offsetting the first window analysis was applied to the fourth correct fault prediction, results show that the model was able to predict a fault between 26 and 36 ms (depending on the offset) prior to the onset.

## 7. Future Work

The work and results described in this paper represent an important first step to realizing a cavity fault prediction system, and also highlight areas where future efforts should be focused to make it a reality. Briefly, they are:

- Implement scheduled training to maintain model performance over time. The RF signals from cavities can, and do, change over time. Whether it is changes to the control software, modifications to the hardware itself, or changing environmental conditions, data drift presents a real problem in large, complex systems such as CEBAF. To effectively handle this, the model requires frequent, scheduled training with recent data to adapt to changes.
- Minimize model inference time. Implementing Monte-Carlo dropout to get a confidence for each prediction requires passing the input through the model multiple times. This may prove to be prohibitively costly for a system that needs to respond on the order of a few hundred milliseconds. Therefore, a combination of optimizing model computational efficiency, but also running the model closer to the source of data is required. There is a general push to port machine learning models to field-programmable gate arrays (FPGA), for example, at a number of accelerator facilities [11, 26-28].
- Perform a test on live streaming data. Absent a major overhaul, in their current as-built state, CEBAF cavities are unable to continuously stream signals. However, a test on live data is an important next step towards bringing the system to fruition. While our model is CEBAF – and even cryomodule – specific, the approach we have described could be adapted to develop a model at a facility where streaming data is available.

- Dialogue with RF subject matter experts and explore strategies for preemptively avoiding a fault and the timescales required to execute them.

## 8. Summary

This work developed a deep learning binary classifier for predicting slow-developing accelerating cavity faults, often hundreds of milliseconds in advance, while minimizing false positives in highly imbalanced datasets. The network architecture integrates LSTM and CNN layers to extract temporal and spatial features from input multivariate time series signals. The model distinguishes between 100 ms samples of pre-fault and normal data. In the context of large and complex systems like CEBAF, minimizing unnecessary changes is prioritized over fault accuracy. Therefore, fault confidence thresholds were individually adjusted for each cavity to minimize false positives. Additionally, a criterion was implemented requiring three consecutive windows to be identified as faulty before a decision is made. We assessed the optimized model's performance on an imbalanced dataset to simulate deployment, demonstrating its ability to correctly identify 4 out of 5 slow developing faults, with three predictions made over 1 second in advance. The model achieved nearly perfect accuracy on normal events (51,993 out of 52,000). The seven misclassified events showed anomalous behavior in the input data (see Appendix A).

These results offer a significant accomplishment in forecasting a specific type of accelerating cavity fault that causes substantial downtime in CEBAF. Currently, CEBAF cavities cannot continuously stream RF signals due to their as-built state, and therefore this system cannot be deployed. Nonetheless, this work serves as a crucial proof-of-concept outcome, emphasizing the necessity of integrating this capability into future system designs. The capacity to predict faults prior to their onset reveals new research avenues for devising preemptive strategies to prevent such faults.

## Acknowledgements

This work is supported by the U.S. Department of Energy, Office of Science, Office of Nuclear Physics under Contract No. DE-AC05-06OR23177. The authors gratefully acknowledge Tom Powers for his diligent work in analyzing and labeling the data.

# Appendix A: Example of a False Positive in the Deployed Dataset

For the test on the deployed dataset, the binary classifier misclassified seven of the 6,500 normal events in cavity 3 as faulty. Figure A1 displays the four RF signals from one of those misclassified events. For context, Fig. A2 displays the signals from the timestamp immediately before and after the event. The yellow highlighted regions in Fig. A1 show the three consecutive windows the model identified as faulty. Comparing those regions with the signals immediately before and after shows that there is clearly anomalous behavior. Even though the larger magnitude spikes did not lead to a fault – and therefore we consider it a misclassification – it is clear why the model interpreted this as portending a fault. It should be noted that if this were a real deployed system, and if it turned out that this cavity was misclassifying too many normal events, the fault confidence threshold could easily be adjusted on the fly to achieve the desired false positive rate.

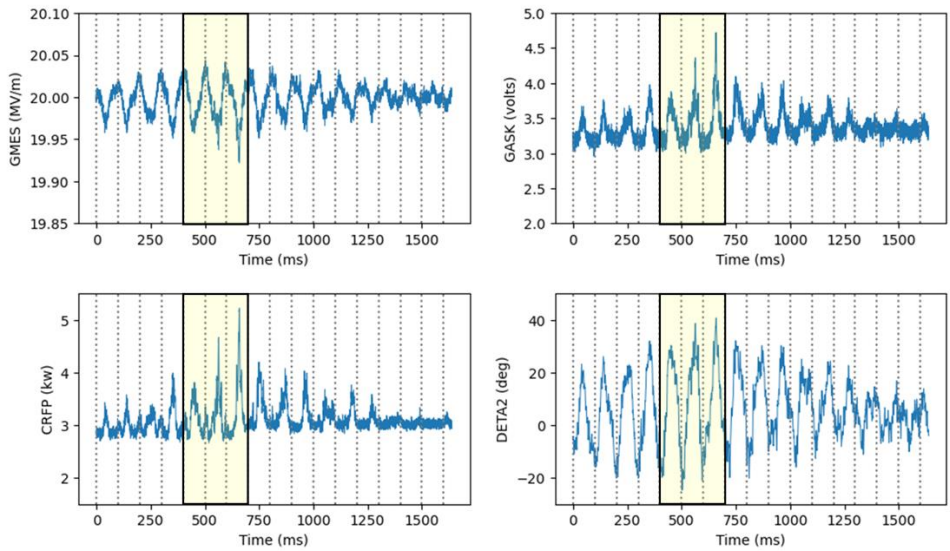

FIGURE A1: Signals from a normal event in cavity 3 which the model identified as a fault. Vertical dashed lines denote the start and end of each 100 ms window and the yellow highlighted areas show the three consecutive windows the model classified as portending a fault.

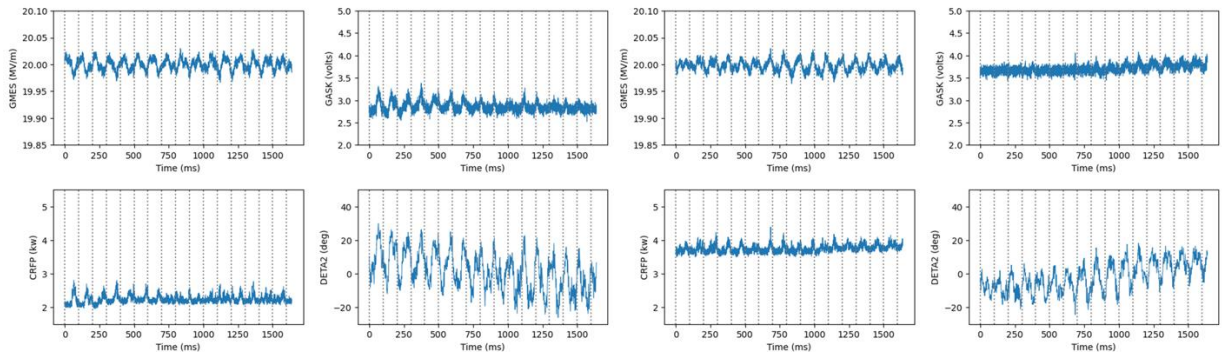

FIGURE A2: Cavity 3 signals collected immediately before (left) and after (right) the event captured in Fig. A1. Vertical dashed lines denote the start and end of each 100 ms window.